\documentclass[a4paper,12pt]{article}
\usepackage{graphicx}
\begin{document}

\title{\Large{\bf Intermittent chaos in a model of financial markets with heterogeneous agents}}

\author{Taisei Kaizoji \\ Division of Social Sciences, International Christian University, \\ 3-10-2 Osawa, Mitaka, Tokyo 181-8585, Japan. \\ and \\
Econophysics Laboratory \\ 
5-9-7-B Higasi-cho, Koganei-shi, Tokyo 184-0011, Japan.}

\date{}
\maketitle 

\begin{abstract}
In this paper we study the price dynamics in a simple model of financial markets with heterogeneous agents. We concentrate on how increases in the total number of active traders influences fluctuations of asset prices. We find that a curious route to chaos is observed when the total number of [active traders] increases. Particularly, we show that {\it intermittent chaos} [1] of price fluctuations is observed as the total number of trader increases. \par
\end{abstract}

\section{Introduction}
In past decades, irregular dynamics of asset prices such as stock prices, bond prices and exchange rates have been a matter of robust debate among academics as well as practitioners. There is already a growing literature attempting to model the heterogeneity in traders' trading strategies that may lead to market instability and complicated dynamics, such as cycles or even chaotic fluctuations in financial markets [2-10]. \par 
One of the most important features in financial time series data is that price fluctuations are characterized by phases of low volatility with small price changes and phases of high volatility with large price changes. A possible and convincing explanation for this phenomenon is that {\it intermittency} of asset price fluctuation is caused by the interaction between heterogeneous traders having different trading strategies and expectations about future prices.\par
Our previous works [11, 12] proposed a discrete-time heterogeneous-agent model of a financial market that explains so-called speculative bubbles. This model showed that when the non-linearity of the traders' excess demand function is sufficiently strong, chaos in asset prices is observed. 
In this paper we will investigate further the dynamical properties of fluctuation of asset price in the heterogeneous-agent model presented in our previous studies. Our model is composed of two-dimensional difference equations. We focus attention on the effect of changes of the total number of active traders on the price dynamics. A number of empirical studies on financial time series have considered the statistical properties of price fluctuations, and variation in the total number of traders intervening in the price adjustment process has been shown by them to be a very realistic focus of interest. We imagine a simple financial market as consisting of a market maker who mediates a risky asset, and two types of traders who can be observed in financial practice\footnote{These two different classes of traders are often used in the finance literature on  heterogeneous-agent systems.}: \textit{fundamentalists} who believe that the asset price will return to the fundamental price, and \textit{chartists} who base their trading decisions on an analysis of past price trends. \par
We find that a curious route to {\it intermittent chaos} is observed when the total number of traders participating in active trading increases. More concretely, the following bifurcation scenario is observed as the number of traders increases: 
\begin{enumerate}
\item when the total number of traders who participate in the market is few, the price converges to the fundamental price. 
\item As the total number of traders increases, first the {\it intermittent chaos} of price fluctuation appears through the {\it period doubling} cascade. 
\item Second, infinitely many {\it period halving} bifurcations occur, and the price fluctuation becomes more regular again.  
\item Third, chaotic price fluctuation appears once again through the {\it period doubling} cascade and the price fluctuations diverge to infinity.
\end{enumerate} 
The plan of the paper is the following: Section 2 presents the heterogeneous-agent model briefly. Section 3 investigates the properties of price dynamics. Section 4 sums up briefly.

\section{The Model}
Let us consider the market for a risky asset, composed of two groups of traders having different trading strategies: {\it fundamentalists} and {\it chartists}. Fundamentalists are assumed to have a reasonable knowledge of the fundamental value of the risky asset. The fundamentalist's strategy can be described as follows: if the price $ p_{t} $ is below the fundamental value $ p^* $, then the fundamentalist tries to buy the risky asset because it is undervalued; if the price $ p_t $ is above the fundamental value $ p^* $, then he tries to sell the risky asset because it is overvalued. The excess demand for the risky asset is given by:
\begin{equation}
x^f_{t} = \exp(\alpha (p^* - p_{t})) - 1,  \alpha > 0,
\end{equation}
where $ \alpha $ is the parameter that denotes the strength of the non-linearity of the fundamentalist excess demand function (1). The fundamentalist's excess demand function (1) is derived in a one-period utility optimizing framework. The technical details of the derivation of (1) within a utility maximizing framework are given in Kaizoji (2001a). We can see that Equation (1) has captured the distinctive features of the fundamentalist's strategy. \par
While fundamentalists calculate the fundamental value, chartists estimate a trend in the price change. Chartists can be assumed to form their expectation of the price of the risky asset according to the simple adaptive scheme: 
\begin{equation}
p^e_{t+1} = p^e_{t} + \mu (p_{t} - p^e_{t}),\quad 0 < \mu \geq 1, 
\end{equation}
where $ p^e_{t} $ denotes the price at period $ t $ expected by chartists, and the parameter $ \mu $ is the error correction coefficient. \par
As above, the chartist's excess demand function is given by: 
\begin{equation}
x^c_{t} = \exp(\beta (p^e_{t+1} - p_{t})) - 1,  \beta > 0,
\end{equation}
where $ \beta $ is the parameter that denotes the strength of the non-linearity of the fundamentalist excess demand function (1). The chartist's excess demand function (3) means that chartists try to buy the risky asset when they anticipate that the price will rise within the next period; inversely, that they try to sell the risky asset when they expect the risky asset price to fall within the next period. \par
Let us now consider the adjustment process of the price in the market. We assume the existence of a market-maker who mediates the trading. 
If the excess demand in period $ t $ is positive (negative), the market maker raises (reduces) the price for the next period $t+1$. 
Let $\kappa $ be the fraction of chartists in the total number of traders. Then the process of price adjustment can be written as 
\begin{equation}
p_{t+1} - p_{t} = \theta N [(1 - \kappa) x^f_{t} + \kappa x^c_{t}],  
\end{equation}
where $\theta $ denotes the speed of the adjustment of the price, and $ N $ the total number of traders. 

\section{Price dynamics} 
Substituting (1), (2) and (3) to (4), the dynamical system can be obtained as 
\begin{eqnarray}
p_{t+1} - p_{t} &=& \theta N [(1 - \kappa) (\exp(\alpha (p^* - p_t)) - 1) \nonumber\\
&+& \kappa (\exp(\beta (1 - \mu) (p^e_{t} - p_t)) - 1)] \nonumber\\
p^e_{t+1} - p^e_{t} &=& \mu (p_t - p^e_t). 
\end{eqnarray}
In the following discussion we highlight the impact of increases in the total number of traders $ n $ on the price fluctuation that is defined as the price increment, $r_{t} = p_{t+1} - p_{t}$. \par 
We first restrict ourselves to investigating the following set of parameters: 
\begin{equation}
\alpha = 3, \quad \beta = 1, \quad \mu = 0.5, \quad \kappa = 0.5, \quad \theta = 0.001. 
\end{equation}
It is clear that the two-dimensional map (5) has a unique equilibrium with $ p^e_t = p_t = p^* $, namely $ (\bar{p}^e, \bar{p}) = (p^*, p^*) $, given the above conditions. \par 
Elementary computations show that for our map (5) the sufficient condition for the local stability of the fixed point $ p^* $ is given as 
\begin{equation}
N < \frac{2 (2 - \mu)}{\theta [\alpha (2 - \mu)(1 - \kappa) 
+ 2 \beta (1 - \mu) \kappa]}.
\end{equation}
From (7) it follows that, starting from a small number of traders $ N $ inside the stability region, when the number of traders $ N $ increases, a loss of stability may occur via a {\it flip bifurcation}. \par
We shall now look more globally into the effect of increases in the number of traders on the price dynamics. Figure 1 shows a bifurcation diagram of the price increments $ r_t $ with $ N $ as the bifurcation parameter under the set of parameters (6). For the convenience of illustration, Figure 1 is drawn using $ \theta N $ as the bifurcation parameter. This figure suggests the following bifurcation scenario. The price increments $ r_t $ converge to $ 0 $ when the number of traders $ N $ is small. In other words, the price converges to the fundamental price $ p^* $ when the active traders are few. However the price dynamics become unstable when the number of traders $ N $ exceeds about $ 1000 $, and chaotic behavior of the price increments occurs after infinitely many period-doubling bifurcations. If $ N $ is further increased, then the price increments $ r_t $ become more regular again after infinitely many period-halving bifurcations\footnote{Period halving bifurcation is observed in other economic models. For examples see [13, 14].}. A stable 2 orbit occurs for an interval of $ N $-values. However as $ N $ is further increased, the behavior of the price increment $ r_t $ becomes once again chaotic, and the prices diverge. \par
Let us investigate closely the characteristics of chaos that are observed in the parameter interval ($ 2000 < N < 4000 $). Figure 2 shows a series of price increments $ r_t $ with $ N = 4000 $ and the set of parameters (6). The figure shows apparently the characteristic of intermittent chaos, that is, a long laminar phase, where the price fluctuations behave regularly, is interrupted from time to time by chaotic bursts\footnote{Intermittency is one of the typical routes of transition from a periodic state to chaos. For more detail on intermittency see Berge, et. al. [1].}. 

\section{Concluding Remarks}
In this paper we propose a simple model of a financial market with heterogeneous traders, that is, fundamentalists and chartists. The result of the computer experiment was that increases in the number of active traders cause chaotic behavior in asset prices. In particular, we show that the chaotic behavior is characterized by intermittency, which means the irregular switching between phases of low volatility and phases of high volatility. \par
In the last decade international financial markets have been repeatedly disturbed by a series of severe financial and currency crises, so that increasing instability of financial markets has attracted the attention of academics and policy makers. 
Our results suggest that rapid increases in the number of agents who participate in financial trade due to globalization of financial trading might be one reason that international financial markets have become more vulnerable. 

\section{Acknowledgements}
Financial support by JSPS is gratefully acknowledged. All remaining errors, of course, are mine.

\newpage
\begin{figure}
 \begin{center}
\end{center}
\caption{The bifurcation diagram. Figure 1 shows a bifurcation diagram of the price increments with $ \theta N $ as the bifurcation parameter. The parameter values: $ \alpha = 3 $, $ \beta = 1 $, $ \mu = 0.5 $, $ \kappa = 0.5 $ and $ \theta = 0.001 $.} 
\label{fig1}
\end{figure}

\begin{figure}
\begin{center}
\end{center}
\caption{Intermittent chaos. Figure 2 shows the time series of the price increments $ r_t $ for the parameter values: $ N = 4000 $, $ \alpha = 3 $, $ \beta = 1 $, $ \mu = 0.5 $, $ \kappa = 0.5 $ and $ \theta = 0.001 $.} 
\label{fig2}
\end{figure}

\end{document}